
\tolerance=10000
\magnification=1200
\raggedbottom

\baselineskip=15pt
\parskip=1\jot

\def\sk{\vskip 3\jot}

\def\heading#1{\vskip3\jot{\noindent\bf #1}}
\def\label#1{{\noindent\it #1}}


\def\ref#1;#2;#3;#4;#5.{\item{[#1]} #2,#3,{\it #4},#5.}
\def\refinbook#1;#2;#3;#4;#5;#6.{\item{[#1]} #2, #3, #4, {\it #5},#6.} 
\def\refbook#1;#2;#3;#4.{\item{[#1]} #2,{\it #3},#4.}


\def\({\bigl(}
\def\){\bigr)}

\def\[{\big[}
\def\]{\big]}

\def\al{\alpha}
\def\be{\beta}
\def\ga{\gamma}
\def\de{\delta}
\def\io{\iota}
\def\la{\lambda}
\def\me{\omega}
\def\ep{\epsilon}

\def\bfb{{\bf b}}
\def\bfv{{\bf v}}
\def\bfw{{\bf w}}
\def\bfx{{\bf x}}
\def\bfr{{\bf r}}
\def\bfs{{\bf s}}

\def\np{\vfill\eject}

\def\cru{commutative ring with unit}
\def\nbyn{{n\times n}}
\def\kbyk{{k\times k}}
\def\kmobykmo{{(k-1)\times (k-1)}}
\def\moton{(-1)^n}

\def\subkbyk{_{k\times k}}
\def\subnbyn{_{n\times n}}

\def\Tr{{\rm Tr}}
\def\maoo{-A_{1,1}}

\def\pmu{p(\mu)}

\def\coeff#1{\langle #1\rangle}

\def\mu{\xi}

{
\pageno=0
\nopagenumbers
\rightline{\tt ffd.tex}
\vskip1in

\centerline{\bf A Formula for the Determinant}
\vskip0.5in

\centerline{Nicholas Pippenger}
\centerline{\tt njp@math.hmc.edu}
\sk

\centerline{Department of Mathematics}
\centerline{Harvey Mudd College}
\centerline{301 Platt Boulevard}
\centerline{Claremont, CA 91711}
\vskip0.5in

\noindent{\bf Abstract:}
We give a formula for the determinant of an $\nbyn$ matrix with entries from a \cru{}.
The formula can be evaluated by a ``straight-line program'' performing only additions, subtractions and multiplications of ring elements; in particular it requires no divisions or conditional branching (as are required, for example, by Gaussian elimination).
The number of operations performed is bounded by a fixed power of $n$, specifically $O(n^4\log n)$. 
Furthermore, the operations can be partitioned into ``stages'' in such a way that the operands of the operations in a given stage are either matrix entries or the results of operations in earlier stages, and the number of stages is bounded by a fixed power of the logarithm of $n$, specifically $O\((\log n)^2\)$.
\sk

\vfill\eject
}

\heading{1. Introduction}
\sk

Let $R$ be a \cru, and let $A\in R^\nbyn$ be an $\nbyn$ matrix with entries in $R$.
The {\it determinant\/} $\det(A)$ of $A$ can be defined in various ways.
It can, for example, be defined abstractly as the unique alternating multilinear function $\det: R^\nbyn\to R$ of the columns of its argument that assumes
the value $1\in R$ for the identity matrix $I\in R^\nbyn$.
(The term ``alternating'' here means that $\det(A)$ vanishes if any two columns of $A$ are equal, and thus that it is multiplied by $-1$ if any two columns are exchanged.)
With this definition, one must of course prove that there is a unique function satisfying these conditions.
On the other hand, it can be defined concretely by the formula
$$\det(A) = \sum_\pi (-1)^{\io(\pi)} \prod_{1\le i\le n} A_{i,\pi(i)}. \eqno(1.1)$$
(The sum here is over all permutations $\pi$ of $\{1,\ldots,n\}$, and $\io(\pi)$ denotes the number of ``inversions'' in $\pi$:
the number of pairs $1\le i<j\le n$ such that $\pi(i)>\pi(j)$.)
Although (1.1) defines the determinant in terms of addition, subtraction and multiplication (operations that are available in any \cru), its direct evaluation requires $n! - 1$ additions and $n!\,(n-1)$ multiplications. 
By exploiting the distributivity of multiplication over addition, the number of multiplications can be reduced to $O(n!)$ (though it does not reduce the number of additions). 
In any case, direct evaluation of (1.1) is infeasible for any but small values of $n$.

If $R$ is a field (so that division by any non-zero element is available), then {\it Gaussian elimination\/} provides a more efficient algorithm for evaluating the determinant.
One performs "row operations" (exchanging two rows, or adding a scalar multiple of one row to another) to transform the matrix to one that is "upper triangular", so that all entries below the main diagonal vanish (that is, so that $A_{i,j} = 0$ for $i>j$).
Since exchanging two rows of a matrix multiplies its determinant by $-1$, and adding a scalar multiple of one row to another leaves its determinant unchanged, the determinant of the original matrix bears a known relationship to that of the transformed matrix.
But the determinant of an upper triangular matrix is simply the product of its diagonal entries (because only one term in (1.1)
is non-zero).
Since each row operation requires $O(n)$ scalar operations, and $O(n^2)$ row operations are required to create ${n\choose 2} = O(n^2)$ zeros below the main diagonal, Gaussian elimination requires just $O(n^3)$ scalar operations, a great improvement over direct evaluation of (1.1).
But this improvement brings with it some disadvantages.
As we have noted, $R$ must be a field, because determining which scalar multiple of one row to add to another requires
division by an element on the main diagonal.
And we may have to perform an exchange of rows to ensure that an element on the main diagonal is non-zero (if no such exchange is possible, the determinant is zero).
This means that besides additions, subtractions, multiplications and divisions, we must also perform tests to determine whether or not an entry is zero, branching accordingly.
Thus, unlike direct evaluation of (1.1), which yields a ``straight-line program'', Gaussian elimination yields a ``branching program''.
The recurring need to perform tests to determine what operation to perform next greatly reduces the opportunities to perform operations ``in parallel''.
To assess these opportunities, we consider partitioning the operations into ``stages'' in such a way that the operations in each stage have operands that are either matrix entries or results computed in earlier stages (so that all operations in a given stage can be performed in parallel).
For Gaussian elimination, the $O(n^3)$ operations can be partitioned into $n$ stages, each comprising $O(n^2)$ operations.
Each stage except for the lastbegins by performing an exchange of rows, if necessary, to get a non-zero element on the main diagonal, if  possible (or to determine that the determinant is zero).
But the divisions in each stage depend on results computed in the immediately preceding stage, to that the $n$ stages must be performed sequentially.

Another class of efficient methods for evaluating the determinant when $R$ is a field goes under the name {\it condensation}.
The term ``condensation'' is often taken to refer to a method introduced by Dodgson [D] in 1866, but Eves [E] describes a simpler method introduced by in 1853.
In either of these methods, the evaluation of an $\nbyn$ determinant is ``condensed'' to that of an $(n-1)\times (n-1)$ determinant; this reduction is repeated until a $1\times 1$ matrix, whose sole entry is its determinant, is obtained. 
Let $A\in R^\nbyn$ be an $\nbyn$ matrix with entries in a field $R$.
Then if $A_{1,1}\not=0$, we have
$$\det(A) = \det(B)/(A_{1,1})^{n-2},$$
where $B\in R^{(n-1)\times(n-1)}$ is the $(n-1)\times(n-1)$ matrix with entries 
$$B_{i,j} = \det\pmatrix{A_{1,1} &A_{1,j} \cr A_{i,1} &A_{i,j}},$$
for $2\le i,j\le n$.
(See Eves [E], pp.~129--130, for a proof of this identity.)
As with Gaussian elimination, we need $R$ to be a field to divide by $(A_{1,1})^{n-2}$ when it is not zero, and we need to perform tests and conditional row (or column) exchanges when it is zero.
While all the entries of $B$ can be computed in parallel using $O(n^2)$ operations, the computation requires $n-1$ stages (each of which begins with a test, and ends by reducing the dimension of the matrix by one), for a total of $O(n^3)$ operations.
(We have presented Chi\'{o}'s version of condensation rather than Dodgson's, because with Chi\'{o}'s, at each stage there is just one matrix entry whose vanishing can cause problems, whereas with Dodgson's there are many.)

Our goal in this paper is to present a formula for the determinant that combines the following three desiderata.
\medskip

\item{(I)}
Like the formula (1.1), it requires only additions, subtractions and multiplications
(so that it can be used for matrices over any \cru).

\item{(II)}
The total number of scalar operations is bounded by a fixed power of $n$.

\item{(III)}
The number of stages of operations is bounded by a fixed power of the logarithm of $n$.
\medskip

\noindent
To avoid prolonging suspense, we state our formula:
$$\det(A) = \moton\,\langle \xi^n\rangle \,  \sum_{0\le i\le n} 
\left(1 - \prod_{1\le k\le n} \sum_{0\le j\le n} \left(( A_{\kbyk})^j\right)_{k,k}\,\xi^j\right)^i, \eqno(1.2)$$
where the expression $\langle \xi^n\rangle \, p(\xi)$ denotes the coefficient of $\xi^n$ in the polynomial $p(\xi)$,
$A_{\kbyk}$ denotes the upper-left $\kbyk$ submatrix of $A$ (with entries $A_{i,j}$ for $1\le i,j\le k$), and
$\left((A_{\kbyk})^j\right)_{k,k}$ denotes the lower-right $(k,k)$-entry of the $j$-th power of $A_{\kbyk}$.
It is clear that condition (I) is met.
We shall see in Section 6 that condition (II) can be met  with a bound of $O(n^4\log n)$ on the number of operations, and condition (III) can be met with a bound of $O\((\log n)^2\)$ on the number of stages.
We obtained formula (1.2) by simplifying an algorithm given by Chistov [Ch]; discussion of Chistov's algorithm, and to earlier work addressing conditions (II) and (III), will be given in Section 7.

Before proceeding, we should mention another way of looking at our problem.
We may introduce $n^2$ indeterminates $\al_{i,j}$ (for $1\le i,j\le n$),  work in the ring 
$R[\al] = R[\al_{1,1},\ldots,\al_{n,n}]$, and seek the determinant $\det\(A(\al)\)$ of the matrix $A(\al)$ whose entries $A(\al)_{i,j}$ are the indeterminates $\al_{i,j}$ for $1\le i,j\le n$.
Our problem then becomes one of {\it constructing\/} the polynomial $\det\(A(\al)\)$ (as an element of $R[\al]$) from the $\al_{i,j}$ and elements of $R$, rather then {\it computing\/} $\det(A)$ (as an element of $R$) from the $A_{i,j}$ and other elements of $R$.
Equation (1.1) then becomes
$$\det\(A(\al)\) = \sum_\pi (-1)^{\io(\pi)} \prod_{1\le i\le n} \al_{i,\pi(i)}.$$
This formulation makes it clear what operations are allowed (namely, ring operations), and that no testing or conditional branching is allowed.
For the sake of simplicity, however, we shall continue to work in the original formulation, except where this alternative adds clarity.
\np

\heading{2. Polynomials}
\sk

Our formula can be modified to give not only  the determinant of $A$, but the entire
{\it characteristic polynomial},
$$c_A(\la) = \det(\la I - A). \eqno(2.1)$$
Here we are taking the determinant of the matrix $\la I - A$, which has entries in the ring of polynomials $R[\la]$.
This determinant is a polynomial of degree $n$, and this polynomial is ``monic'' (that is, the coefficient of the leading term is $1$), since that term arises, as the product of the entries on the main diagonal, from the identity permutation in (1.1).
Thus we have
$$c(\la) = \la^n + c_{n-1}\,\la^{n-1} + \cdots + c_1\,\la + c_0.$$
(Though $c_A(\la)$ depends on $A$, we shall take $A$ to be a fixed matrix throughout our discussion, and drop the subscript $A$.)
Setting $\la = 0$ in (2.1) gives  
$$c(0) = \det(-A) = \moton\,\det(A),$$
so we have
$$\eqalign{
\det(A)
&= \moton\,c(0) \cr
&= \moton\,c_0 \cr
&= \moton\,\langle \la^0\rangle \, c(\la). \cr
}$$

We shall actually find it more convenient to compute the ``reversal'' of the characteristic polynomial,
$$d(\mu) = \mu^n \, c(\mu^{-1}). \eqno(2.2)$$
This is simply the characteristic polynomial with the order of its coefficients reversed; that is,
$$d(\mu) = d_n\,\mu^n + d_{n-1}\,\mu^{n-1} + \cdots + d_1\,\mu + 1,$$
where $d_k = c_{n-k}$ for $1\le k\le n$.
As  $c(\la)$ is monic, $d(\mu)$ is what we shall call, for the nonce, ``unic'', meaning that its constant term is $1$.
We then have
$$\det(A) = \moton\,\langle \mu^n\rangle \, d(\mu). \eqno(2.3)$$
\np

\heading{3. Formal Power Series}
\sk

Let $S$ be a commutative ring with unit, and let $S[\mu]$ be the ring of polynomials with coefficients in $S$, which is also a commutative ring with unit.
An embarrassing deficiency of polynomials in $S[\mu]$ is that few of them have multiplicative inverses: only the elements of $S$ with multiplicative inverses have multiplicative inverses in $S[\mu]$.
We shall ameliorate this deficiency somewhat by embedding the ring $S[\mu]$ of polynomials in the ring $S[[\mu]]$ of ``formal power series''.
A formal power series is like a polynomial, but with a potentially infinite number of terms:
$$p(\mu) = p_0 + p_1\,\mu + p_2\,\mu^2 + \cdots + p_k\,\mu^k + \cdots\,.$$
These objects can be added and multiplied ($r(\mu) = p(\mu)+q(\mu)$ means $r_k = p_k + q_k$ and 
$r(\mu) = p(\mu)\,q(\mu)$ means $r_k = p_0\,q_k + p_1\,q_{k-1} + \cdots + p_{k-1}\,q_1 + p_k\,q_0$, just as for polynomials), and they form a \cru{} under these operations.
(See Niven [N] for more information about formal power series.)

The term ``formal'' means that we do not contemplate substituting a non-zero element of $S$ for $\mu$ and inquiring whether the  infinite sum on the right-hand side converges to some element of $S$; indeed, there may not be any notion of convergence in $S$.
(We can of course substitute zero of $\mu$ and obtain the constant term: $p(0) = p_0$.)
There is, however, a useful notion of convergence in $S[[\mu]]$.
Let $(\mu^k)$ denote the ideal comprising all elements of $S[[\mu]]$ that are multiples of $\mu^k$ (so $p(\mu)\in (\mu^k)$ if and only if $p_0 = \cdots = p_{k-1} = 0$).
Then we shall say that a sequence $p^1(\mu),p^2(\mu),\ldots,p^l(\mu)$ converges to the limit $q(\mu)$ when and only when, for every $k\ge 1$, $q(\mu) - p^l(\mu) \in (\mu^k)$ for all sufficiently large $l\ge 1$
(that is, when more and more terms of $p^l(\mu)$ agree with those of $q(\mu)$ as $l$ increases).
Addition and multiplication in $S[[\mu]]$ become continuous operations under this notion of convergence.

We can now observe that $p(\mu)$ has a multiplicative inverse $p(\mu)^{-1}$ whenever $p_0$ has a multiplicative inverse.
We first note that when $p(\mu)$ is unic (that is, when $p_0=1$), we can write $p(\mu) = 1 - q(\mu)$ with $q(\mu)\in(\mu)$; we then have the formula
$$p(\mu)^{-1} = 1 + q(\mu) + q(\mu)^2 + \cdots + q(\mu)^k + \cdots\,.  \eqno(3.1)$$
To see this, we note that
$$\(1 - q(\mu)\) \, \(1 + q(\mu) + \cdots + q(\mu)^k\) = 1 - q(\mu)^{k+1},$$
so
$$\(1 - q(\mu)\) \, \(1 + q(\mu) + \cdots + q(\mu)^k\) - 1 \in (\mu^{k+1}),$$
which means that the series on the right-hand side of (3.1) converges to an element of $S[[\mu]]$ that, when multiplied by $p(\mu) = 1-q(\mu)$, yields $1$.
If $p(\mu)$ is not unic, but $p_0$ has a multiplicative inverse, then $p_0^{-1}\,p(\mu)$ is unic, and we have
$p(\mu)^{-1} = p_0^{-1} \,\(p_0^{-1} \, p(\mu)\)^{-1}$.

Of course, formal power series have infinitely many coefficients, and we cannot compute them all.
Nor do we need to:  the first $k$ coefficients of the reciprocal of a unic formal power series $p(\mu)$ depend only on the first $k$ coefficients of $p(\mu)$, and similarly for the coefficients of the sum, difference and product of two such series.
Since we are ultimately interested in a polynomial $d(\mu)$ of degree at most $n$, we may simply pretend that all coefficients of monomials $\mu^k$ for $k\ge n+1$ vanish when reciprocating or multiplying unic power series.
Another way of looking at this is to consider the quotient ring $S[[\mu]]/(\mu^{n+1})$ of $S[[\mu]]$ divided by the ideal 
$(\mu^{n+1})$ generated by monomial $\mu^{n+1}$.
Each element of this quotient ring is an equivalence class that contains exactly one polynomial of degree at most $n$.
This has the effect of endowing the module $S[\mu]_n$ of polynomials of degree at most $n$ with a new multiplication operation, whereby it becomes a ring in which each unic polynomial $p(\mu) = 1 - q(\mu)\in S[\mu]_n$ has a multiplicative inverse 
$p(\mu)^{-1}\in S[\mu]_n$ given by
$$p(\mu)^{-1} = \sum_{0\le j\le n} q(\mu)^j. \eqno(3.2)$$
\sk

\heading{4. The Inverse of a Matrix}
\sk

We shall need one more tool to derive our formula for the determinant, and that is a formula for
the multiplicative inverse $C^{-1}\in S^\kbyk$ of a matrix $C\in S^\kbyk$, when $\det(C)$ has a multiplicative inverse in $S$.
The $(i,j)$-th entry of the inverse is
$$(C^{-1})_{i,j} = (-1)^{i+j} \, \det(C[j,i]) \,\det(C)^{-1}, \eqno(4.1)$$
where the matrix $C[j,i] \in S^\kmobykmo$ is obtained from $C$ by deleting all entries in either the $j$-th row or the $i$-th column, or both.
The matrix $(-1)^{i+j} \, \det(C[j,i])$ is called the ``adjugate'' (or sometimes the ``adjoint'') of $A$.
(See Eves [E], pp.~153--154 for a proof of (4.1).)
Formula (4.1) is equivalent to finding each column $\bfx$ of $C^{-1}$ by using Cramer's rule (see Eves [E], p.~157) to 
solve the equation $C\bfx = \bfb$, where $\bfb$ is the corresponding column of $I$.

Suppose $B$ is an $\nbyn$ matrix with entries in a commutative ring with unit $S$.
For $1\le k\le n$, let $ B_{\kbyk}$ denote the $\kbyk$ submatrix of $B$ comprising those entries in its first $k$ rows and in its first $k$ columns.
Then from (4.1) we have 
$$\((B_{\kbyk})^{-1}\)_{k,k} = \det(B_{(k-1)\times(k-1)})\,\det(B_{\kbyk})^{-1}, \eqno(4.2)$$
because $(-1)^{k+k} = 1$, and deleting the last row and column from $B_{\kbyk}$ yields $B_{(k-1)\times(k-1)}$.

Multiplying (4.2) for $1\le k\le n$ yields
$$\eqalignno{
\prod_{1\le k\le n} \((B\subkbyk)^{-1}\)_{k,k}
&= \det(B_{0\times 0}) \, \det(B\subnbyn)^{-1} \cr
&= \det(B)^{-1}, &(4.3)\cr
}$$
where we have used $\det(B_{0\times 0}) = 1$ (the determinant of a $0\times 0$ matrix has, according to (1.1), $0! = 1$ term, which is a product of zero factors, and is therefore the multiplicative identity $1$).
Inverting and exchanging the two sides of (4.3), we obtain
$$\det(B) = \left(\prod_{1\le k\le n} \((B\subkbyk)^{-1}\)_{k,k}\right)^{-1}. \eqno(4.4)$$
\sk

\heading{5. The Formula}
\sk

Taking $B = I - \mu A$ in (4.4), we obtain
$$\eqalignno{
d(\mu)
&=
\det(I - \mu A) \cr
&= \left(\prod_{1\le k\le n} \(\((I - \mu A)\subkbyk\)^{-1}\)_{k,k}\right)^{-1}. &(5.1) \cr
}$$
We shall apply the reciprocation formula (3.2) to the two reciprocations that appear in (5.1).

The inner reciprocation in (5.1) is of $I\subkbyk - \mu A\subkbyk$, which is a unic element of $R^\kbyk[\mu]_n$. 
The ring $R^\kbyk$, however, is not commutative if $k\ge 2$, so we cannot take $S = R^\kbyk$.
To use (3.2), we must find a commutative subring $S\subseteq R^\kbyk$ that contains all the quantities appearing in (3.2).
Accordingly, we shall take $S$ to be the
subring of $R^\kbyk$ comprising linear combinations, with coefficients from $R$, of powers of $A\subkbyk$; 
this $S$ is a \cru.
(If, for example, we take $A$ to be the matrix $A(\al)$ of indeterminates $\al_{i,j}$, then $A(\al)^2$ is a matrix whose entries are quadratic forms in the $\al_{i,j}$, $A(\al)^3$ is a matrix whose entries are cubic forms in the $\al_{i,j}$,
and so forth.)
Applying (3.2), we obtain
$$d(\mu) = 
\left(\prod_{1\le k\le n} \left(\sum_{0\le j\le n} (\mu\,A\subkbyk)^j\right)_{\!\!k,k}\right)^{-1}.$$
Since taking the $(k,k)$-th entry of a matrix is a linear operation, this identity becomes
$$d(\mu) = 
\left(\prod_{1\le k\le n} \sum_{0\le j\le n} \left((A\subkbyk)^j\right)_{k,k}\,\mu^j\right)^{-1}. \eqno(5.2)$$

Each factor in the product in (5.2) is unic (because the term for $j=0$ is $((A\subkbyk)^0)_{k,k} = (I\subkbyk)_{k,k} = 1$), so the product itself is unic.
Thus the outer reciprocation in (5.1) is of this product in $R[\mu]_n$.
If $p(\mu)$ is a unic formal power series, we can take $q(\mu) = 1 - p(\mu)$ in (3.2), obtaining
$$\eqalign{
p(\mu)^{-1}
&= \sum_{i\ge 0} \(1 - p(\mu)\)^i. \cr
}$$
Thus (5.2) becomes
$$d(\mu) = \sum_{0\le i\le n} 
\left(1 - \prod_{1\le k\le n} \sum_{0\le j\le n} \left((A\subkbyk)^j\right)_{k,k}\,\mu^j\right)^i. $$
Finally, using (2.3) we obtain (1.2).
Alternatively, using (2.2) we can obtain the complete characteristic polynomial,
$$c(\la) = \sum_{0\le m\le n} \la^m \, \langle \mu^{n-m}\rangle   \sum_{0\le i\le n} 
\left(1 - \prod_{1\le k\le n} \sum_{0\le j\le n} \left((A\subkbyk)^j\right)_{k,k}\,\mu^j\right)^i.$$

It is amusing to see how (1.2) works when $A$ is the $1\times 1$ matrix with sole entry $\al$.
The inner sum (with two terms) in (1.2) is then $1 + \al\mu$; the product (with one factor) is also $1 + \al\mu$;
the outer sum (with two terms) is then $1 - \al\mu$; taking $-\coeff{\mu}$, the determinant is 
$\al$.
(We have performed the polynomial multiplication in $R[\mu]_1$, dropping terms in the ideal $(\mu^2)$.)
Less trivially, for the $2\times 2$ matrix $A = \pmatrix{\al &\be \cr \ga &\de}$, the inner sum for $k = 1$ 
(with three terms) is $1 + \al\mu + \al^2\mu^2$, the inner sum for $k = 2$ (with three terms) is
$1 + \de\mu + (\be\ga + \de^2)\mu^2$; the product (with two factors) is
$1 + (\al + \de)\mu + (\al^2 + \al\de + \be\de + \de^2)\mu^2$; the outer sum (with three terms) is
$1 - (\al + \de)\mu + (\al\de - \be\ga)\mu^2$; taking $\coeff{\mu^2}$, the determinant is $\al\de - \be\ga$ (and taking $-\coeff{\mu}$, the trace is $\al + \de$).
(We have performed the polynomial multiplications in $R[\mu]_2$, dropping terms in the ideal $(\mu^3)$.)
\np

\heading{6. Evaluating the Formula}
\sk

We shall start assuming that multiplications of $n\times n$ matrices are performed by evaluating $n^2$ inner products in parallel, and that inner products are performed using $O(n)$ operations in $O(\log n)$ stages.
(To see that $O(\log n)$ stages suffice to sum the $n$ terms of an inner product, we exploit the associative law of addition:
break the terms into pairs and add the terms in each pair, then break the resulting sums into pairs and add the results in each pair, and so forth.)
Thus we shall assume that matrix multiplications are performed using $O(n^3)$ operations in $O(\log n)$ stages.
(Faster matrix multiplication methods will be discussed later.)

We shall also assume that multiplication of polynomials $R[\xi]_n$ is performed by evaluating $n+1$ coefficients in parallel, and that each coefficient is evaluated by summing $O(n)$ terms in $O(\log n)$ stages
(again exploiting the associativity of addition).
Thus we shall assume that polynomial multiplications are performed using $O(n^2)$ operations in $O(\log n)$ stages. 
(Faster polynomial multiplication methods will be discussed later.)

The evaluation of (1.2) can be divided into two parts: (A) evaluation of the coefficients
$\left((A\subkbyk)^j\right)_{k,k}$ for $0\le k\le n$ and $0\le j\le n$ (this will be dominated by matrix multiplications, and will 
give the main contribution to the number of operations), and (B) evaluation of the coefficients of $d(\mu)$ from the results of part (A) (this will be dominated by polynomial multiplication).

We begin by considering part (B), assuming part (A) to have already been completed.
Apart from some additions and subtractions of polynomials, we must (1) multiply the $n$ polynomials that are factors 
in the product in (5.3), and (2) compute the $n+1$ powers of polynomials that are the terms in the outer sum of (5.3).
Part (1) calls for $n-1$ polynomial multiplications, reach requiring $O(n^2)$ operations in $O(\log n)$ stages.
They can all be performed using $O(n^3)$ operations in $O\((\log n)^2\)$ stages, this time by exploiting the associative law for polynomial multiplication: break the factors into pairs, multiply the factors in each pair, then break the resulting products into pairs and multiply the products in each pair, and so forth; there are $O(\log n)$ stages of polynomial multiplication,
and therefore $O\((\log n)^2\)$ stages of operations.

Part (2) of part (B) calls for the computation of the powers $p(\mu), p(\mu)^2,\ldots,p(\mu)^n$ of a polynomial $p(\mu)$.
This can be effected by $n-1$ polynomial multiplications, each using $O(n^2)$ operations, for a total of $O(n^3)$ operations.
We shall show how these operations can be performed in $O\((\log n)^2\)$ stages.
We start by squaring $p(\mu)$ to obtain $p(\mu)^2$, and we then multiply $p(\mu)$ by $p(\mu)^2$ to obtain $p(\mu)^3$.
Next we square $p(\mu)^2$ to obtain $p(\mu)^4$, and we the multiply $\pmu$, $\pmu^2$ and $\pmu^3$ by $\pmu^4$ (in parallel) to obtain $\pmu^5$, $\pmu^6$ and $\pmu^7$, respectively.
We continue in this way, roughly doubling the number of powers of $\pmu$ that we have computed in each ``round'', and thus obtaining all $n$ powers in $O(\log n)$ rounds.
Each polynomial multiplication produces a new power of $\pmu$, so there are still $n-1$ polynomial multiplications for a total of $O(n^3)$ operations.
But now there are  $O(\log n)$ rounds, each containing just two stages of polynomial multiplication and therefore $O\((\log n)^2\)$ stages of operations.
Thus part (B) can be performed with $O(n^3)$ operations in $O\((\log n)^2\)$ stages.

 Part (A) calls for the evaluation of  $\left((A\subkbyk)^j\right)_{k,k}$ for $1\le k\le n$ and $0\le j\le n$.
 It will suffice to consider the case of $k = n$; the other values of $k$ involve the same operations on smaller matrices, and 
 can be performed in parallel with the operations for $k = n$; they can therefore be accounted for by including an extra factor of $n$ in the bound on the number of operations for $k = n$.
 Thus we want to evaluate $(A^j)_{n,n}$ for $0\le j\le n$.
 We could compute the powers $A, A^2,\ldots, A^n$ in the same way we computed the powers $\pmu, \pmu^2,\ldots,\pmu^n$ in part (2) of part (B) (just replacing polynomial multiplications by matrix multiplications), then select the $(n,n)$-th entry from each power.
 This would use $n-1$ matrix multiplications in $O(\log n)$ stages of matrix multiplication, or $O(n^4)$ operations in
 $O\((\log n)^2\)$ stages of operations; including an extra factor of $n$ to account for the $n$ values of $k$ would give 
 $O(n^5)$ operations in $O\((\log n)^2\)$ stages of operations
 
 But it seems wasteful to compute an $\nbyn$ matrix just to select a single entry; we shall show how to reduce the number of operations to $O(n^4\log n)$, while maintaining the number of stages at $O\((\log n)^2\)$.
 We have $(A^j)_{n,n} = \bfv A^j \bfw$, where $\bfv$ is the $1\times n$ matrix $\pmatrix{0 &\cdots &0 &1}$ and $\bfw$ is the $n\times 1$ transpose of $\bfv$.
 The product $\bfv A^j$ is just the last row of the matrix $A^j$, so we shall just compute the last rows of the powers of $A$.
 We start by squaring $A$ to obtain $A^2$, then multiply the $1\times n$ matrix whose row is $\bfv A$ 
 on the right by $A^2$ to obtain the $1\times n$ matrix whose row is $\bfv A^3$.
 Next we square $A^2$ to obtain $A^4$, then multiply the $2\times n$ matrix whose rows are $\bfv A$, $\bfv A^2$ and $\bfv A^3$
 on the right by $A^4$ to obtain the $3\times n$ matrix whose rows are $\bfv A^5$, $\bfv A^6$ and $\bfv A^7$.
 Continuing in this way, we roughly double the number of rows $\bfv A^j$ we have computed in each ``round'', and we obtain all $n$ rows $\bfv A, \bfv A^2,\ldots, \bfv A^n$ in just $O(\log n)$ rounds, using just $O(\log n)$ matrix multiplications, or $O(n^3\log n)$ operations in $O\((\log n)^2\)$ stages.
 Multiplying these rows on the right by $\bfw$ amounts to selecting the rightmost entry from each row, and entails no further computation.
Including an extra factor of $n$ to account for the $n$ values of $k$  gives 
 $O(n^4\log n)$ operations in $O\((\log n)^2\)$ stages of operations.
 
 In our analysis, we have assumed that matrix multiplication is performed in the naive way, using $O(n^3)$ operations.
It has been known that matrix multiplication can be performed using $O(n^\me)$ operations for some $\me<3$
since 1969, when Strassen [S1] obtained $\me = \log_2 7 < 2.81$.
Further improvements have been obtained since then, but these improved results may depend on the existence of certain constants in the ring $R$.
Strassen's original result, however, requires only the constants $\pm 1$, present in any ring with unit.

We have also assumed that multiplication of polynomials of degree at most $n$ is performed in the naive way, using
$O(n^2)$ operations.
Here, too, improvements are available (under the name ``fast Fourier transforms''), yielding bounds as good as 
$O(n\log n)$ in some cases.
But these results also depend on the existence of certain constants (usually ``roots of unity''), and in any case polynomial multiplication does not make the dominant contribution to the number of operations for our formula.
\sk

\heading{7. Some History}

The formula we have presented is the culmination of a long line of research aimed at achieving desiderata (II) and (III) for matrices with entries in a field $K$.
The first algorithm satisfying (II) and (III) was presented by Csanky [Cs] in 1976.
He begins with the observation that the trace $\Tr(A)$ of $A$ is the sum $\la_1 + \cdots + \la_n$ of the eigenvalues of $A$.
Since the eigenvalues of $A^k$ are the $k$-th powers of the eigenvalues of $A$, we have
$\Tr(A^k) = \la_1^k + \cdots + \la_n^k$, the ``power-sum symmetric function'' of the eigenvalues.
(The eigenvalues might belong only to an extension of the field $K$ containing the matrix entries, but symmetric functions of them all belong to $K$.)
The coefficients of the characteristic polynomial $c_A(\la)$ are the ``elementary symmetric functions'', and they can be obtained from the power-sum symmetric function by solving a triangular system of linear equations:
$$\eqalign{
\pmatrix{
1 &0 &0 &\cdots &0 \cr
p_1 &2 &0 &\cdots &0 \cr
p_2 &p_1 &3 &\cdots &0 \cr
\vdots &\vdots &\vdots &\ddots &\vdots \cr
p_{n-1} &p_{n-2} &p_{n-2} &\cdots &n \cr}
\pmatrix{c_{n-1} \cr c_{n-2} \cr c_{n-3} \cr \vdots \cr c_0 \cr}
= - \pmatrix{p_1 \cr p_2 \cr p_3 \cr \vdots \cr p_n \cr}.
}$$
(The equations in this system are sometimes called ``Newton's identities'' for symmetric functions.)
Csanky shows how these computations can be performed with $O(n^4)$ operations in $O\((\log n)^2\)$ stages, but the presence of the constants $1,\ldots, n$ on the main diagonal means that the triangular system is singular if any of these constants vanishes in the field $K$; thus the algorithm requires that the characteristic of $K$ be either $0$, or a prime $p>n$.

Berkowitz [Be], in 1984, was the first to give an algorithm achieving (II) and (III), without performing any divisions, for any field $K$; since it performs only ring operations, it also satisfies (I).
He begins by proving that if the matrix has the form
$$A = \pmatrix{A_{1,1} &\bfr \cr \bfs &B},$$
where $\bfr$ is $1\times (n-1)$, $\bfs$ is $(n-1)\times 1$ and $B$ is $(n-1)\times (n-1)$, then the coefficients of 
$$c_A(\la) = c_n \,\la^n + c_{n-1}\,\la^{n-1} + \cdots + c_1\,\la + c_0$$ 
are related to those of 
$$c_B(\la) = c'_{n-1}\,\la^{n-1} + \cdots + c'_1\,\la + c'_0$$ 
by multiplying the vector of the latter by an $(n+1)\times n$ matrix:
$$\pmatrix{c_n \cr c_{n-1} \cr c_{n-2} \cr c_{n-3} \cr \vdots \cr c_0} = 
\pmatrix{1 &0 &0 &\cdots &0 \cr \maoo &1 &0 &\cdots &0 \cr -\bfr\bfs &\maoo &1 &\cdots &0 \cr
-\bfr B\bfs &-\bfr\bfs &\maoo &\cdots &0 \cr \vdots &\vdots &\vdots &&\vdots \cr 
-\bfr B^{n-2}\bfs &-\bfr B^{n-3}\bfs &-\bfr B^{n-4}\bfs &\cdots &\maoo \cr}
\pmatrix{c'_{n-1} \cr c'_{n-2} \cr c'_{n-3} \cr \vdots \cr c'_0}.$$
A product of $n-1$ such matrices thus relates the coefficients of $c_A(\la)$ to the coefficient $-A_{n,n}$ of the characteristic polynomial of the lower-right $1\times 1$ submatrix of $A$.
The verification of Berkowitz's algorithm relies on algebraic identities of considerable complexity, the combinatorial interpretation of which has been explored by Mahajan and Vinay [M1, M2].
An advantage of Berkowitz's algorithm is that it has been adapted to the computation of ``Pfaffians'' by Mahajan, Subramanya and Vinay [M3].
(The determinant of a skew-symmetric matrix (a matrix whose transpose is its negative) is the square of a polynomial in the matrix entries; the square-root of the determinant (with appropriate sign) is called the Pfaffian.)
Berkowitz claims a bound of $O(n^{4+\ep})$ operations, for any $\ep>0$, without using fast matrix multiplication; with care this bound can be reduced to $O(n^4\log n)$, just as in this paper.
The number of stages is also $O\((\log n)^2\)$, as in this paper.
Finding various powers of various submatrices of $A$ at heart of both Berkowitz's algorithm and ours, so the number of operations for both should be similar.
But the actual number of operations for Berkowitz's algorithm is definitely larger: he computes an inner product where we select the $(k,k)$-th entry of a matrix, and he performs the $n-1$ matrix multiplications mentioned above where we reciprocate a unic polynomial in $R[\mu]_n$.

As we mentioned in the introduction, the algorithm we present is a simplification of one described in 1985 by Chistov [Ch], who assumed matrix entries from a field, used an iterative method for the outer reciprocation, and gave cruder bounds for the number of operations.
It should be mentioned that as early as 1982 it was observed by Borodin, von zur Gathen and Hopcroft [Bo] that desiderata (II) and (III) could be obtained, without performing any divisions, in the weak form of $O(n^{15})$ operations and 
$O\((\log n)^2\)$ stages, by combining a very general result of Strassen [S2] on elimination of division with a very general result of Valiant, Skyum, Berkowitz and Rackoff [V] on reduction of the number of stages.
\sk

\heading{References}
\sk
\baselineskip13pt

\ref Be; S. J. Berkowitz;
``On Computing the Determinant in Small Parallel Time Using a Small Number of Processors'';
Inform.\ Process.\ Lett.; 18 (1984) 147--150.

\ref Bo; A. Borodin, J. von zur Gathen and J. Hopcroft;
``Fast Parallel Matrix and GCD Computations'';
Inform.\  Control; 52 (1982) 241--256.

\refinbook Ch; A. L. Chistov;
``Fast Parallel Calculation of the Rank of Matrices over a Field of Arbitrary Characteristic'';
in: L.~Budach (editor); Fundamentals of Computation Theory; Springer Lecture Notes in Computer Science, 199 (1985) 63--69.


\ref Cs; L. Csanky;
``Fast Parallel Matrix Inversion Algorithms'';
SIAM J. Comput.; 5:4 (1976) 618--623.

\ref D; C. L. Dodgson;
``Condensation of Determinants'';
Proc.\ Roy.\ Soc.\ London; 15 (1866--1867) 150--155.

\refbook E; H. W. Eves;
Elementary Matrix Theory; Allyn and Bacon, 1966; Dover, 1980.

\ref M1; M. Mahajan and V. Vinay;
``Determinant: Combinatorics, Algorithms, and Complexity'';
Chicago J. Theor.\ Comput.\ Sci.; 1997:5.

\ref M2; M. Mahajan and V. Vinay;
``Determinant: Old Algorithms, New Insights'';
SIAM J.\ Discrete Math.; 12:4 (1999) 474--490.

\ref M3; M. Mahajan, P. R. Subramanya and V. Vinay;
``The Combinatorial Approach Yields an NC Algorithm for Computing Pfaffians'';
Discrete Appl.\ Math.; 143:1--3 (2004) 1--16.

\ref N; I. Niven;
``Formal Power Series'';
Amer.\ Math.\  Monthly; 76:8 (1969) 871--889.

\ref S1; V. Strassen;
``Gaussian Elimination Is Not Optimal'';
Num.\ Math.; 13 (1969) 354--356.

\ref S2; V. Strassen;
``Vermeidung von Divisionen'';
J.\ reine angew.\ Math.; 264 (1973) 184--202.

\ref V; L. G. Valiant, S. Skyum, S. Berkowitz and C. Rackoff;
``Fast Parallel Computation of Polynomials Using Few Processors'';
SIAM J. Comput.; 12:4 (1983) 641--644.


\bye